# Gender Equity in Physics in India: Interventions, Outcomes, and Roadmap

Srubabati Goswami[1, a)], Aru Beri[2], Bindu Bambah[3], Deepa Chari[4], V. Madhurima[5], Gautam Menon[6], Vandana Nanal[7], Pragya Pandey[1], Tanusri Saha-Dasgupta[8], Aditi Sen-De[9], and Prajval Shastri[10]

[1]*Physical Research Laboratory, Ahmedabad, India;* [2]*Indian Institute of Science Education and Research, Mohali;* [3]*University of Hyderabad, Telangana;* [4]*Homi Bhabha Centre for Science Education, TIFR, Mumbai;* [5]*Central University of Tamil Nadu, Thiruvarur;* [6]*Ashoka University, Sonepat;* [7]*DNAP, Tata Institute of Fundamental Research, Mumbai;* [8]*CMPMS, S. N. Bose Centre, Kolkata;* [9]*Harish-Chandra Research Institute, Prayagraj;* [10]*International Centre for Theoretical Sciences, Bengaluru*

a)sruba.goswami@gmail.com

**Abstract.** The gender imbalance in physics higher education and advanced professions is a global problem, and India is not an exception. Although the issue has been acknowledged widely, discrimination needs to be recognized as the driving force. The past three years have witnessed initiatives by different gender groups as well as the Government of India in addressing these lacunae. We report various activities, describe interventions, and present statistics indicating improvements achieved. The Gender in Physics Working Group has brought about significant gender reforms in the Indian Physics Association. The working group organized an open discussion on the issue of sexual harassment in physics professions for the first time in 2018. Subsequently, in 2019, GIPWG organized the first-ever national conference on gender issues, Pressing for Progress. The deliberations of the conference culminated in the Hyderabad Charter, a roadmap towards gender equity in India. The Working Group for Gender Equity constituted under the Astronomical Society of India, also played an impactful role. At the government level, notable new initiatives include Gender Advancement through Transforming Institutions and the proposed Science and Technology Innovation Policy for mainstreaming equity and inclusion.

## THE GENDER DIVIDE IN PHYSICS

The science, technology, engineering, and math disciplines in India in general, and physics in particular, reveal a glaring gender gap. To achieve a quantitative assessment, a gender survey was conducted in 2021 by the Gender in Physics Working Group (GIPWG) of the Indian Physics Association (IPA), among select universities and institutes. The data show a gross underrepresentation of women as we go up the ladder. The survey indicated that the percentage of women research scholars in physics is about 31%, for postdoctoral fellows it is about 29%, and for faculty the percentage drops to about 14%, demonstrating the leaky pipeline effect (panel 1 of Fig. 1). The number plummets further when one considers awards and accolades. When looking at the total number of faculty members, prominent "male peaks" appear for the elite research/teaching institutes (panel 2 of Fig. 1). The highest Indian honor in science, the coveted Shanti Swarup Bhatnagar Award, which started in 1901, so far has seen only one female recipient in physics, in 2018. The percentage of women in prestigious science academies stands at ~6%. A 2017 minisurvey of women in leadership positions such as directors, vice-chancellors, or members of a governing council, revealed a meager 3% representation of women [1]. This number, however, showed a promising increase to 16% in 2021 (panel 3 of Fig. 1).

The establishment of the GIPWG as a part of the IPA has demonstrated visible overall improvements (panels 4–6 of Fig. 1). The percentage of women authors in *Physics News,* the IPA newsletter, has steadily increased over the last three years. The percentage of women in the IPA's executive committee is ~30%. Since 2017, the number of women who have received the IPA award has gone up from 9% to 23%, and the number of female speakers in the prestigious DAE C. V. Raman Lecture Series has increased from 8% to 24%.

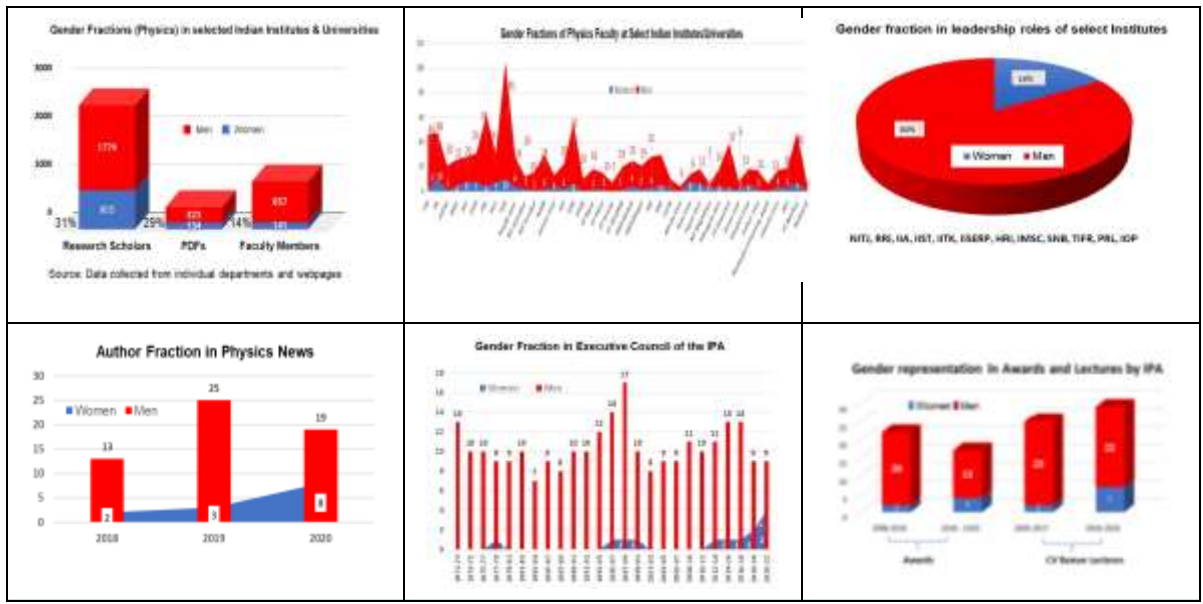

**FIGURE 1.** Gender percentages in various aspects of the physics profession.

## INITIATIVES, INTERVENTIONS, AND OUTCOME

The period 2017–2020 saw continued initiatives coming from government agencies and national science academies. Two physics-specific gender groups, GIPWG (under IPA) [2] and the Working Group on Gender Equity (WGGE, under the Astronomical Society of India) [3], also played effective roles.

The GIPWG was launched in 2017 with Professor Prajval Shastri as the chair. The group seeks to promote gender equity in the Indian physics profession and organized two Pressing for Progress (PFP) events in the last three years:

1. In 2018, PFP served as a discussion meeting on the gender gap in physics at ICTS, Bengaluru. One of the highlights of this event was an open discussion on sexual harassment in academia, a first for India.
2. In 2019 [4], PFP was the first interdisciplinary conference in physics at the University of Hyderabad. About 250 physicists, social scientists, educators, and diversity practitioners attended the conference to deliberate on the issue of gender diversity in physics. The program included physics keynote talks, invited talks and posters, and a panel discussion titled "Gender Gap in Physics: Whose Problem Is It?" In an interdisciplinary plenary session, social scientists and diversity experts explored different avenues for promoting gender equity. Additionally, there were three interactive workshops to understand different challenges in gender inequity. These deliberations and discourses culminated in the Hyderabad Charter for Gender Equity in Physics [5], which advocates evidence-based interventions that step away from the patriarchal and "fix the women" frameworks. The charter contains recommendations for institutions, physics teaching, conferences, and national agencies. The GIPWG is also facilitating gender working groups in physics subdisciplines.

*Physics News* published a special issue on the first woman high-energy physicist of India, Dr. Bibha Chowdhuri [6]. This issue is aimed at illustrating the isolation and difficulties faced by a woman studying physics and celebrating the life and work of a pioneer to create intergenerational solidarity. Her alma maters, Physical Research Laboratory and Tata Institute of Fundamental Research (TIFR), have now instituted public lectures in her honor.

The WGGE has also played an increasingly impactful role through plenary sessions in the ASI annual meetings dedicated to gender equity that include open discussions, regularizing child care at conferences, regular nationwide talks on gender inequity through the Anna Mani lecture series (named after an eminent female meteorologist), and gender sensitization by interdisciplinary scholars, as well as carrying out gender audits of Indian research institutes.

Other notable efforts include the website BiasWatchIndia.com, which is engaged in documenting women's representation in Indian science conferences and at workshops/meetings, and in highlighting gender-biased panels. TheLifeOfScience.com, an independent media initiative formed in 2016, is involved in popularizing work done by marginalized groups. They published an illustrated book, *31 Fantastic Adventures in Science*, narrating tales of 31 female scientists. Another noteworthy initiative is the Vigyan Vidushi summer school for female students pursuing master's degrees in physics, organized by TIFR [7], one of the leading research institutes in India, with an aim to mentor women at a stage when they are making a career choice.

Government of India agencies initiated several new schemes. The Department of Science and Technology (DST) introduced power grants and power fellowships for women researchers. In 2020, the DST launched an accreditation program, Gender Advancement through Transforming Institutions [8], along the lines of the United Kingdom's Athena Swan Charter. The current year, 2021, also saw the emergence of a landmark document: the proposed Science and Technology Innovation Policy [9]. Highlights of this document include the creation of an equity and inclusion charter, a mandate of at least 30% women in leadership positions, flexible timings, gender-neutral child care benefits, couple recruitment policies, retirement, and spousal benefits for the LGBTQ+ community.

## A ROADMAP FOR THE FUTURE

The past three years have witnessed incremental changes as well as the emergence of radical ideas to address gender inequity. The implementation of the proposed gender equity policies at the institutional level can bring much-needed change in our perception of gender and can create more equitable workspaces. We also need continued efforts to address biases, mandatory gender sensitization programs, nationwide surveys and collection of statistics, and career development and mentoring programs for young students. Involving social scientists in discussions on the best way to achieve gender equality, and men as partners in enabling this transformation, are important directions for the future. An important goal is to step beyond binary gender, and the ultimate goal would be to make gender a nonissue.

## REFERENCES

1. L. Resmi, P. Shastri, S. Goswami, P. Pandey, V. Nanal, P. Kharb *et al.*, "Gender status in the Indian physics profession and the way forward," in AIP Conference Proceedings 2109, edited by B. Cunningham, S. Ghose, and C. O'Riordan (American Institute of Physics, Melville, NY, 2019), 050019. doi.org/10.1063/1.5110093.
2. "The Gender in Physics Working Group (GIPWG) of the IPA," Gender in Physics Working Group (2018), https://www.tifr.res.in/~ipa1970/gipwg/index.php.
3. "Working Group for Gender Equity," Working Group for Gender Equity, (2021), https://astron-soc.in/wgge/about-wgge.
4. P. Shastri, B. Bambah, S. Goswami, and V. Nanal, "A first-of-its-kind national conference towards gender equity by the Indian Physics Association: Pressing for progress 2019," Bulletin of the Association of Asia Pacific Physical Societies, **29**(6), 20–24 (2019).
5. P. Shastri, "The Hyderabad Charter for Gender Equity in Physics 2020," Bulletin of the Association of Asia Pacific Physical Societies **30**(4), 50–52 (2020).
6. *Physics News* **51**(1–2) (2021). https://www.tifr.res.in/~ipa1970/news/2021/JanJune/PN_JanJune21_Index.html.
7. A. Dighe, A. Mazumdar, and V. Nanal, "Vigyan Vidushi 2020," Bulletin of the Association of Asia Pacific Physical Societies **30**(5), 23–24 (2020).
8. "Women Scientists Programs," Department of Science and Technology, Government of India, (18 November, 2021), https://dst.gov.in/scientific-programmes/scientific-engineering-research/women-scientists-programs.
9. "Science, Technology and Innovation Policy 2020," Office of the Principal Scientific Adviser to the Government of India, https://www.psa.gov.in/stip.